\documentclass[a4paper,10pt]{article}
\usepackage{graphicx}
\usepackage{comment}
\usepackage{amssymb}
\usepackage{amsmath}
\usepackage{nicefrac}
\usepackage{graphicx}
\usepackage{dcolumn}
\usepackage{bm}
\usepackage{comment}

\begin{document}

\newcommand{\bin}[2]{\left(\begin{array}{c} \!\!#1\!\! \\  \!\!#2\!\! \end{array}\right)}
\newcommand{\threej}[6]{\left(\begin{array}{ccc} #1 & #2 & #3 \\ #4 & #5 & #6 \end{array}\right)}
\newcommand{\sixj}[6]{\left\{\begin{array}{ccc} #1 & #2 & #3 \\ #4 & #5 & #6 \end{array}\right\}}
\newcommand{\hp}{\mathcal{P}}

\huge

\begin{center}
Regularities and symmetries in atomic structure and spectra
\end{center}

\vspace{0.5cm}

\large

\begin{center}
Jean-Christophe Pain\footnote{jean-christophe.pain@cea.fr}
\end{center}

\normalsize

\begin{center}
\it CEA, DAM, DIF, F-91297 Arpajon, France
\end{center}

\vspace{0.5cm}

\begin{abstract}
The use of statistical methods for the description of complex quantum systems was primarily motivated by the failure of a line-by-line interpretation of atomic spectra. Such methods reveal regularities and trends in the distributions of levels and lines. In the past, much attention was paid to the distribution of energy levels (Wigner surmise, random-matrix model...). However, information about the distribution of the lines (energy and strength) is lacking. Thirty years ago, Learner found empirically an unexpected law: the logarithm of the number of lines whose intensities lie between $2^kI_0$ and $2^{k+1}I_0$, $I_0$ being a reference intensity and $k$ an integer, is a decreasing linear function of $k$. In the present work, the fractal nature of such an intriguing regularity is outlined and a calculation of its fractal dimension is proposed. Other peculiarities are also presented, such as the fact that the distribution of line strengths follows Benford's law of anomalous numbers, the existence of additional selection rules (PH coupling), the symmetry with respect to a quarter of the subshell in the spin-adapted space (LL coupling) and the odd-even staggering in the distribution of quantum numbers, pointed out by Bauche and Coss\'e.
\end{abstract}

\section{Introduction}\label{sec1}

The concept of complex spectra was introduced as a contrast to simple spectra, for which the formation of series is the most conspicuous feature, see Ref. \cite{JOHANSSON96}. Complex spectra are characterized by the appearance of multiplets or groups of lines with characteristic spacings and strengths. Kuhn assigned them to atoms or ions having more than one valence electron (or hole) with $\ell>0$ \cite{KUHN62}. Another distinction between simple and complex spectra, based on the richness in lines and the efforts in unravelling the spectra, was given by Racah at the Rydberg Centennial Conference in Lund in 1954: ``The atomic spectra can be divided in three classes:

(i) spectra where a $p$-subshell is filling up, which may be considered as simple spectra,

(ii) spectra where a $d$-subshell is filling up, which are considered as complex spectra and

(iii) spectra, where a $f$-subshell is filling up, which are so complex that most of them have not been analyzed at all'' \cite{RACAH55}. 

Concerning point (iii), things have changed since then, of course: the analysis of the $4f$-elements (lanthanides or rare earths) improved in the seventies due to better experimental and theoretical tools. In his paper entitled ``Complex Atomic Spectra'', Judd \cite{JUDD85} gives an extensive review of the theoretical treatment of complex atoms and a detailed list of all systems of configurations that had been studied before 1985. Judd also points out the difference between the definition of complex spectra based on the atomic structure and the ``common misconception'' that complex has to do with the complicated line spectrum caused by, \emph{e.g.} overlapping multiplets. One possible way to define the difference between simple and complex spectra could involve the series limits. The basic term system in simple spectra has only one series limit, whereas complex spectra have multiple series limits. This distinction is very close to the one given by Kuhn, while we find both simple and complex spectra in all three classes discussed by Racah. Simple spectra have one common convergence point, the ionization limit, which can be determined by extrapolating regular Rydberg series. The multiple series limits in complex spectra are due to a set of $LS$ terms that may even belong to more than one configuration. Thus, a distinction between simple and complex spectra based on the number of series limits includes the structure of the atomic core, which to a great extent determines the structure of the atom itself. A large number of valence electrons and the competition in binding energy between $d$ and $s$ electrons give rise to a multitude of closely spaced core terms, which act as multiple series limits. 

The link between the distribution of spectral lines and quantum chaos was studied in detail in Refs. \cite{FLAMBAUM94,FLAMBAUM98}. As long as $n$ and $\ell$ remain good quantum numbers, the independent-particle model and the central field approximation both apply, and quantum chaos does not arise. There are two situations where chaos can emerge: the complete breakdown of the independent electron approximation (due, for example, to strong correlations) and a distortion of the central field approximation, due, for instance, to a strong external field. However, the role of chaos in the spectroscopy of highly excited atoms should not be overstated. Due to the Pauli principle, the shell structure of atoms restores spherical symmetry to the many-electron atom at each new row of the periodic table, and spherical symmetry, which helps the independent particle model, inhibits chaos \cite{CONNERADE98}. In addition, as the excitation energy increases, auto-ionization and Auger effects limit the emergence of chaos, because the lifetimes are so short that instabilities in the underlying dynamics do not have time to develop. Situations favourable to the emergence of chaos are those in which a dense manifold of Rydberg states is subject to perturbations with a strength comparable to the level spacing. This constitutes the major difference between chaos and complexity; in some situations the levels do not interact, but are simply very numerous, which makes their analysis impossible in practice although possible in principle. Transition arrays can be described by the UTA (Unresolved Transition Arrays) formalism \cite{BAUCHE79}, which enables one to calculate the first two strength-weighted moments of the distribution of the lines. Such an approach is useful for the interpretation of spectra, but is not related to the problem of quantum chaos.

Cowan compared the term analysis of a spectrum with ``the problem of trying to put together the pieces of a complicated jigsaw puzzle when the pieces never fit exactly, some pieces fit spuriously, some critical pieces are missing and there are pieces present that belong to one or more entirely different puzzles'' \cite{COWAN81}. For that reason, identifying regularities and trends is important for the development of approximate statistical models required in complex spectra due to the huge number of energy levels and electric-dipole lines. Thus, the uncovered regularities and trends enable one to evaluate the reliability of theoretical predictions as well as experimental determinations, and open the way to a better understanding of the underlying physical processes. For instance, let us consider transitions for which principal and orbital (azimuthal) quantum numbers of the upper and lower states are simultaneously changed by unity, \emph{i.e.} transitions of the kind $n\ell\rightarrow n'\ell'$ with $n'=n+1$, $\ell=n-1$ and $\ell'=n'-1=n$ (for instance $1s\rightarrow 2p$, $2p\rightarrow 3d$, $3d\rightarrow 4f$). The hydrogenic expression of the oscillator strength (Gordon's formula) reads \cite{SARANDAEV95,BETHE57}:

\begin{eqnarray}
f_{n\ell\rightarrow n'\ell'}&=&\frac{1}{6}\frac{(2n+1)n^{2n+4}(n+1)^{2n+2}}{(2n-1)(n+1/2)^{4n+6}}\nonumber\\
&=&\frac{1}{3}\left(n+\frac{1}{4(n+1)}\right)\frac{1}{\left(1-\frac{1}{2n}\right)}\frac{1}{\left(1+\frac{1}{4n(n+1)}\right)^{2n+3}}
\end{eqnarray}

\noindent and follows, for large values of $n$, a linear variation

\begin{equation}
f_{n\ell\rightarrow n'\ell'}\approx \frac{n}{3}.
\end{equation}

\noindent The knowledge of such asymptotic behaviour can be useful when preparing experiments. 

In this work, we will first try to explain the fractal nature of some atomic spectra and the importance of scale invariance through Learner's logarithmic rule (in Sec. \ref{sec2}) and the law of anomalous numbers for digits of line strengths (see Sec. \ref{sec3}). Those first two points are not fully understood yet. In Sec. \ref{sec4}, the concept of propensity rule is presented and the possible occurrence of ``blue wings'' in the vicinity of particle-hole coupling is discussed in Sec. \ref{sec5}. Such tendencies can be understood thanks to the selection rules. In Sec. 6 we discuss the symmetry with respect to the quarter of the subshell, which was discovered for the ratio of the intensities of the strongest lines emitted by lanthanide elements in calibrated arcs. Such a symmetry property gives the possibility to interpolate the experimental values more accurately. The studies of hot plasmas have shown the efficiency of the statistical approaches to complex atomic spectra. For instance, approximate formulas have been derived for the distribution of the values of the total angular momentum $J$ in an electronic configuration. However, the existing formulae overlook the fluctuations. One particular fluctuation is the asymmetry in the parity of the $J$ values: there are more even $J$ values than odd ones. We finish therefore, in Sec. \ref{sec7}, with a study of the odd-even staggering in the distribution of total angular momentum $J$. We show that the excess obtained by Bauche and Coss\'e can be obtained in a straightforward and elegant way by generating functions.

\section{Statistics of the lines in intensity octaves: Learner's logarithmic rule}\label{sec2}

\subsection{Phenomenology}\label{subsec21}

Thirty years ago, Learner found what he called an ``unexpected law'' related to the number of weak lines in a spectrum \cite{LEARNER82}. He measured a large number of line intensities in the atomic spectrum of iron, over a dynamic range of 1000 and emphasized the existence of a remarkable power law for the density of lines versus their intensity: the logarithm $\log(N_n)$ of the number of lines whose intensities lie between $2^nI_0$ and $2^{n+1}I_0$ is a decreasing linear function of $n$:
 
\begin{equation}
\log_{10}\left(\frac{N_n}{L}\right)\approx -n.p+a_0, 
\end{equation}
 
\noindent where $L$ is the total number of lines and $p$ the slope. In Learner's example, \emph{i.e.} the spectrum of neutral Iron (Fe I), the value of $I_0$ is chosen in such a way that this law holds for $1\leq n\leq 9$ (9 octaves) when about 1500 lines within the wavelength range 290 nm$\leq \lambda\leq$ 550 nm are considered. One has  $N_n=N_0. 10^{-np}$, where $N_0$ is the number of lines in the first octave: the number of lines is divided by $10^p$ when the size of the interval is multiplied by two. Furthermore, analysis of astrophysical data for arsenic \cite{LEARNER82,SCHEELINE86a,SCHEELINE86b,HOWARD85}  also revealed that a line with the same slope could fit the observed data.

\begin{figure}[ht]
\begin{center}
\vspace{1cm}
\includegraphics[width=10cm]{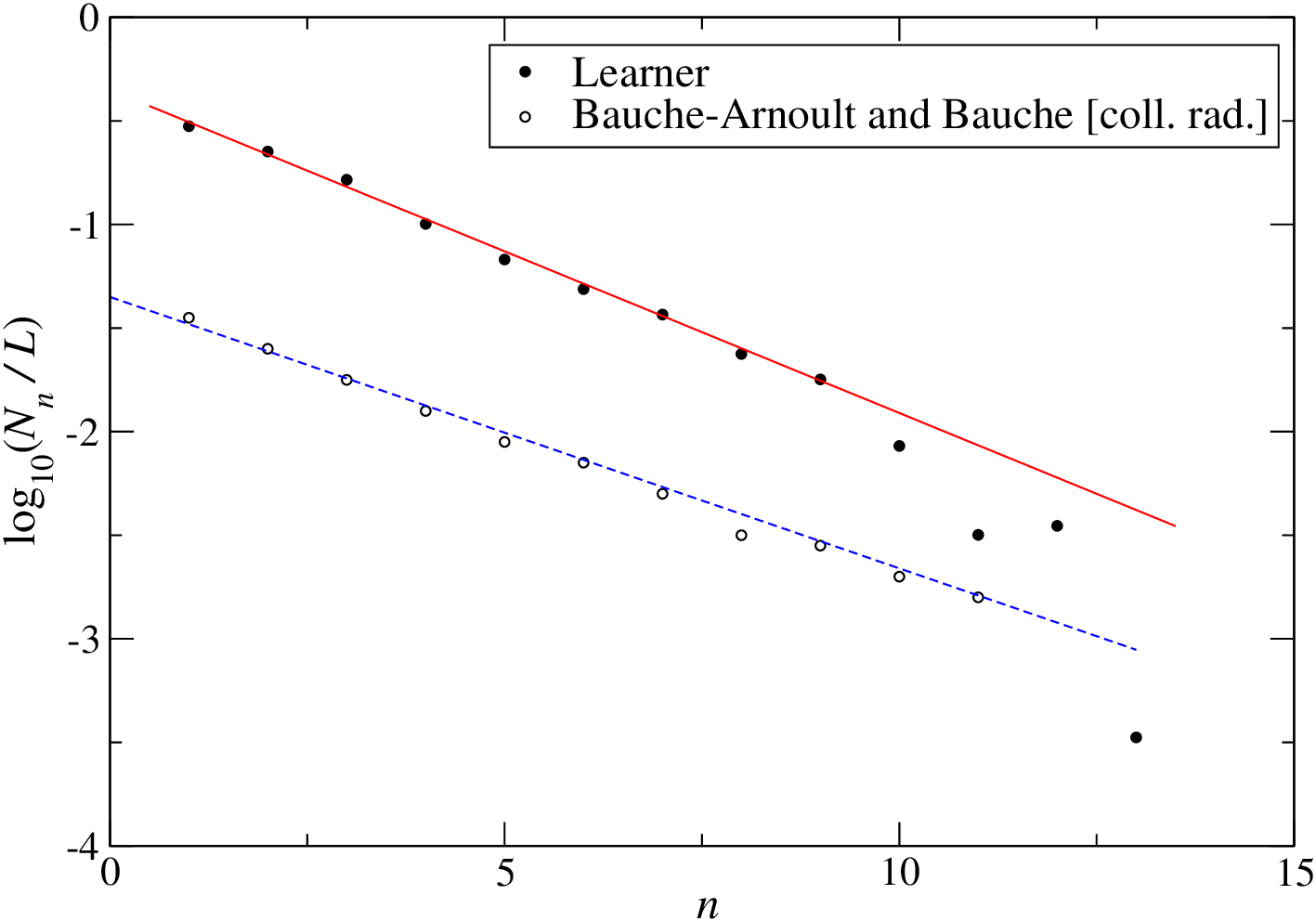}
\end{center}
\vspace{1cm}
\caption{(Color online) Experimental line intensities measured by Learner and values obtained by Bauche-Arnoult and Bauche by means of the collisional-radiative model in the case of the Fe I spectrum.}\label{fig1}
\end{figure}

\begin{table}[t]
\begin{center}
\begin{tabular}{|c|c|c|}\hline
Estimation & $p$ & $D_F$ \\ \hline\hline
Learner & 0.157 & 0.3428 \\ 
Boltzmann populations & 0.118 & 0.2816 \\ 
Col.-rad. populations & 0.131 & 0.3032 \\ 
Kurucz & 0.127 & 0.2967 \\\hline
\end{tabular}
\end{center}
\caption{Value of the fractal dimension $D_F$ for different estimates of the slope $p$.}\label{tab1}
\end{table}

\begin{table}[t]
\begin{center}
\begin{tabular}{|c|c|c|}\hline
Fractal & Dimension \\ \hline\hline
Zeros of the graph of a Brownian function (Wigner's process) & 0.5 \\ 
Feigenbaum's attractor & 0.538 \\ 
Cantor & $\frac{\ln(2)}{\ln(3)}\approx$0.6309 \\ 
Koch curve & $\frac{\ln(4)}{\ln(3)}\approx$1.2619 \\
Julia & 1.2683 \\
Mandelbrot & 2 \\\hline
\end{tabular}
\end{center}
\caption{Dimension of several well-known fractals.}\label{tab2}
\end{table}

Scheeline performed a careful study of Learner's law using hydrogenic electric-dipole transitions. He confirmed the global tendency observed by Learner, but could not provide a precise theoretical interpretation since hydrogenic formulas \cite{SCHEELINE86a} are not relevant for Fe I. Learner, as quoted by Scheeline, also mentioned that, in fields sufficient to cause ionization potential lowering, selection rules may be relaxed so that a large number of forbidden lines in isolated atoms could become allowed. It would be interesting to perform additional calculations (and measurements) to access that suggestion.

Fifteen years later, Bauche-Arnoult and Bauche investigated Learner's results, on the basis of two assumptions for the level populations: Boltzmann equilibrium and collisional-radiative steady-state \cite{BAUCHE97a} (see Fig. \ref{fig1}). In both cases, Learner's linear law is reproduced, but with different slopes, given by 25 \% and 17 \% respectively. The extended line-by-line calculations carried out by Kurucz \cite{KURUCZ} yield a 19 \% difference in the slope, using the assumption of Boltzmann equilibrium. Bauche-Arnoult and Bauche also pointed out the fact that the theoretical work should be improved: more lines should be added to the set of Fe I lines, second-order configuration-mixing effects should be accounted for and a more sophisticated collisional-radiative model should be applied. 

The theory of fractals provides insights into the description of sophisticated shapes of nature. Relying on scale invariance or self-similarity, the shape of a fractal object does not change when the observation scale changes \cite{MANDELBROT82}. The fractal character of the distribution of atomic energy levels was studied by Cederbaum \emph{et al.} \cite{CEDERBAUM85,ZIMMERMANN87,ZIMMERMANN88} and later by Wang and Ong \cite{WANG97}, but very few results exist about the connection between fractals and spectral lines. An application to the spectral lines of hydrogen has been published by DeVito and Little \cite{DEVITO88} who provided, based on an extension of the Cantor set, an interpretation in terms of L\'evy sets \cite{LEVY34} of Balmer's formula giving the energies of the hydrogen atom:

\begin{equation}\label{devito}
E_{ij}=R_{\infty}\left(\frac{1}{i^2}-\frac{1}{j^2}\right),
\end{equation}

\noindent where $R_{\infty}$=13.60569172 eV is the Rydberg constant. The authors used an approach based on the selection of the test function whose L\'evy set (the set of self-similar points) has the same distribution as that of the physical quantity being described. However, their study concerns only the hydrogen atom - to date none has been performed on more complex systems \cite{BAUCHE90}) - and is limited to the energies of the lines, and not the intensities. Moreover, their approach can only be applied to sufficiently simple structures and the L\'evy set yields information only on geometrical properties of the distribution function carrier. More recently, a (multi-) fractal interpretation of fractal properties of the vibrational-rotational absorption bands of water vapor was proposed by Kistenev \emph{et al.} \cite{KISTENEV01}.

\subsection{Distribution for the line strengths}\label{subsec22}

The intensity $I_{ij}$ of a line $i\rightarrow j$ measured by Learner can be expressed as

\begin{equation}\label{emiss}
I_{ij}\propto S_{ij}E_{ij}^4,
\end{equation}

\noindent where $S_{ij}$ and $E_{ij}$ represent the energy and the strength of the line, respectively. Rosenzweig and Porter \cite{ROSENZWEIG60} plotted the empirical distribution of the nearest-neighbor spacings for the odd-parity levels of neutral hafnium and the empirical distribution was found to be well described by the Wigner distribution. Later, Flambaum \emph{et al.} \cite{FLAMBAUM94} studied the spacing distributions of the $J^{\pi}=4^+$ levels of cesium and obtained a good agreement between the experimental level spacings and the Wigner distribution. Porter and Thomas have shown that the amplitudes of the lines between all the levels of two random matrices obey a Gaussian distribution \cite{PORTER56,PORTER65,GRIMES83}, which is 

\begin{equation}\label{port}
D(S)=\frac{L}{\sqrt{2\pi \bar{S}S}}\exp\left(-\frac{S}{2\bar{S}}\right),
\end{equation}

\noindent where $L$ and $\bar{S}$ are the number of lines and the average value of the line strength $S$, respectively, and the strength is defined as the square of the amplitude. The Porter-Thomas law is strictly valid when upper- and lower-level Hamiltonian matrices are random matrices that belong to the Gaussian Orthogonal Ensemble (GOE), which provides a realistic description only inside a $(J,J')$ set. For the whole line set in a genuine transition array (which is actually a large superposition of $(J,J')$ sets \cite{BAUCHE97a}), it is better to use the distribution \cite{BAUCHE91}:

\begin{equation}\label{poissa}
D(S)=\frac{L}{\sqrt{2\bar{S}S}}\exp\left(-\sqrt{2\frac{S}{\bar{S}}}\right).
\end{equation}

\noindent However, there is some numerical evidence of the usefulness of Porter-Thomas law as long as the correlation between the strength of a line and its position within the array is taken into account. In their paper, Flambaum \emph{et al.} compared the line strengths measured in cesium by Bisson \emph{et al.} \cite{BISSON91} with Porter-Thomas distribution. The agreement was good, but the authors refined the results taking into account the minimal threshold intensity that can be experimentally measured, within the modified Porter-Thomas distribution

\begin{equation}
D(S)=\frac{A}{\sqrt{2\pi S\bar{S}}}\exp\left(-\frac{S}{2\bar{S}}\right)\left[1-\left(\frac{S_{\mathrm{min}}}{S}\right)^{1/3}\right],
\end{equation}

\noindent where $A$ is a normalization factor and $S_{\mathrm{min}}$ is the minimal strength that can be observed experimentally. 

\subsection{Distribution $D(S)$ consistent with Learner's rule}\label{subsec23}

Learner's observations can be formulated as

\begin{equation}\label{leap}
\log_{10}\left(\int_{2^{n-1}S_0}^{2^nS_0}D(S)dS\right)\approx -n.p+a_0,
\end{equation}

\noindent where $S_0$ is the reference strength, which is proportional to $I_0$. Searching for a distribution of the form $D(S)\approx k.S^{-\alpha}$, Eq. (\ref{leap}) becomes

\begin{eqnarray}\label{leap2}
\log_{10}\left(\int_{2^{n-1}S_0}^{2^nS_0}D(S)dS\right)&\approx& \log_{10}\left(k\frac{\left[S^{1-\alpha}\right]_{2^{n-1}S_0}^{2^nS_0}}{1-\alpha}\right)\nonumber\\
&\approx&-n.(\alpha-1)\log_{10}(2)+\log_{10}\left(\frac{kS_0^{1-\alpha}\left(1-2^{\alpha-1}\right)}{(1-\alpha)}\right).\nonumber\\
\end{eqnarray}

\noindent which yields $a_0\equiv\log_{10}\left(\frac{kS_0^{1-\alpha}\left(1-2^{\alpha-1}\right)}{(1-\alpha)}\right)$ and $p\equiv(\alpha-1)\log_{10}(2)$. Since $p\approx\frac{1}{2}\log_{10}(2)$, one finds $\alpha=\frac{3}{2}$ and

\begin{equation}\label{dislea}
D(S)\approx k.S^{-3/2}.
\end{equation}

\noindent Such a distribution resembles the Porter-Thomas distribution only that is is a rapidly decreasing function of $S$. It relates to the statistics of the lines of intermingled arrays, but it has not yet been interpreted.
 
\subsection{Fractal dimension of Learner's distribution}\label{subsec24}

The different definitions of the fractal dimension or ways of calculating it (box-counting, Minkowski-Bouligand \cite{BOULIGAND29}, Hausdorff, \emph{etc.}) are described in Ref. \cite{MANDELBROT82}. Suppose that the intensities are distributed regularly in their interval of definition. The fractal dimension can be obtained through the numbering of the $M$ intervals (bars) of size $h$ required to cover the whole ensemble. The fractal dimension $D_F$ is then defined as \cite{KOLMOGOROV58}:

\begin{equation}\label{deffrac}
M=\lim_{h\rightarrow 0}h^{-D_F}.
\end{equation}

\noindent In the interval $\left[2^nI_0,2^{n+1}I_0\right]$ of length $2^nI_0$, the line intensities are separated by a distance

\begin{equation}
d_n=\frac{2^nI_0}{N_0~10^{-np}}=\frac{x^nI_0}{N_0}\;\;\;\;\;\;\;\;\mathrm{with}\;\;\;\;\;\;\;\;x=2~.~10^p.
\end{equation}

\noindent For $h\leq d_n$, one bar is required for each line intensity:

\begin{equation}
M_n= \frac{N_0}{10^{np}}.
\end{equation}

\noindent For $h>d_n$, one has to fill the $n^{th}$ interval of size $2^nI_0$ with bars of size $h$; the number of such bars is 

\begin{equation}
M_n=\frac{2^nI_0}{h}. 
\end{equation}

\noindent Let $n_0$ be the integer such that $h=d_{n_0}$; one finds

\begin{equation}\label{defn0}
n_0=\mathrm{Int}\left[\frac{\ln\left(N_0h\right)-\ln\left(I_0\right)}{\ln(x)}\right],
\end{equation}

\noindent where $\mathrm{Int}$ represents the integer part. The total number of bars of size $h$ required to cover the ensemble is therefore

\begin{equation}
M=\sum_{n=1}^{\infty}M_n=\sum_{n=1}^{n_0}\frac{2^nI_0}{h}+\sum_{n=n_0+1}^{\infty}\frac{N_0}{10^{np}}.
\end{equation}

\noindent Assuming $h<<1$ (in other words $n_0>>1$) enables one to write

\begin{equation}
M\approx\frac{I_0}{h}2^{n_0+1}+\frac{N_0}{10^{\left(n_0+1\right).p}}\frac{1}{\left(1-10^{-p}\right)}.
\end{equation}

\noindent The first term dominates in the latter expression. Using Eq. (\ref{defn0}) for $n_0$ yields

\begin{equation}
M\approx\frac{2}{h}\left(\frac{N_0h}{I_0}\right)^{\frac{\ln 2}{\ln\left(2.10^p\right)}},
\end{equation}

\noindent which leads to the fractal dimension
 
\begin{equation}
D_F=1-\frac{\ln 2}{\ln\left(2.10^p\right)}=1-\frac{\log_{10}(2)}{\log_{10}(2)+p}=\frac{p}{\log_{10}(2)+p}.
\end{equation}

\noindent Learner noticed that $p\approx\log_{10}(2)/2\approx 0.1505$; in that case

\begin{equation}
D_F=\frac{\log_{10}(2)/2}{\log_{10}(2)+\log_{10}(2)/2}=\frac{1}{3}.
\end{equation}

\noindent The values of $D_F$ for the slopes observed by Learner and obtained by Bauche-Arnoult and Bauche are displayed in Table \ref{tab1} and compared to usual values encountered in well-known fractals \cite{MANDELBROT82} (see Table \ref{tab2}). The fact that they are quite small means that the fractal has a low space-filling capacity. 

\section{The law of anomalous numbers}\label{sec3}

Learner's rule concerns the values of the line strengths. It is also worth investigating the digits involved in such numerical values.

\subsection{Presentation}\label{subsec31}

At the end of the nineteenth century, Newcomb \cite{NEWCOMB81} observed that the first pages of logarithm books were more used than the last ones, which led him to conjecture that the significant digits of many sets of naturally occurring data are not equi-probably distributed, but in a way that favors smaller significant digits. For instance, the first significant digit, \emph{i.e.}, the first digit which is non zero, see Table \ref{tab3}, will be 6 more frequently than 7 and the first three significant digits will be 439 more often than 462. Such a regularity was observed for example in mountain heights or river lengths. In the USA, income tax agencies use software based on Benford's law to detect tax fraud. If the data of a tax return more or less fit the law, they are most probably honest. Benford proposed a probability distribution function for significant digits, which states that the probability that the first significant digit $d_1$ is equal to $k$ is given by \cite{BENFORD38}:

\begin{equation}\label{bl2}
\hp\left(d_1=k\right)=\log_{10}\left(1+\frac{1}{k}\right). 
\end{equation}

\begin{figure}
\begin{center}
\vspace{1cm}
\includegraphics[width=10cm]{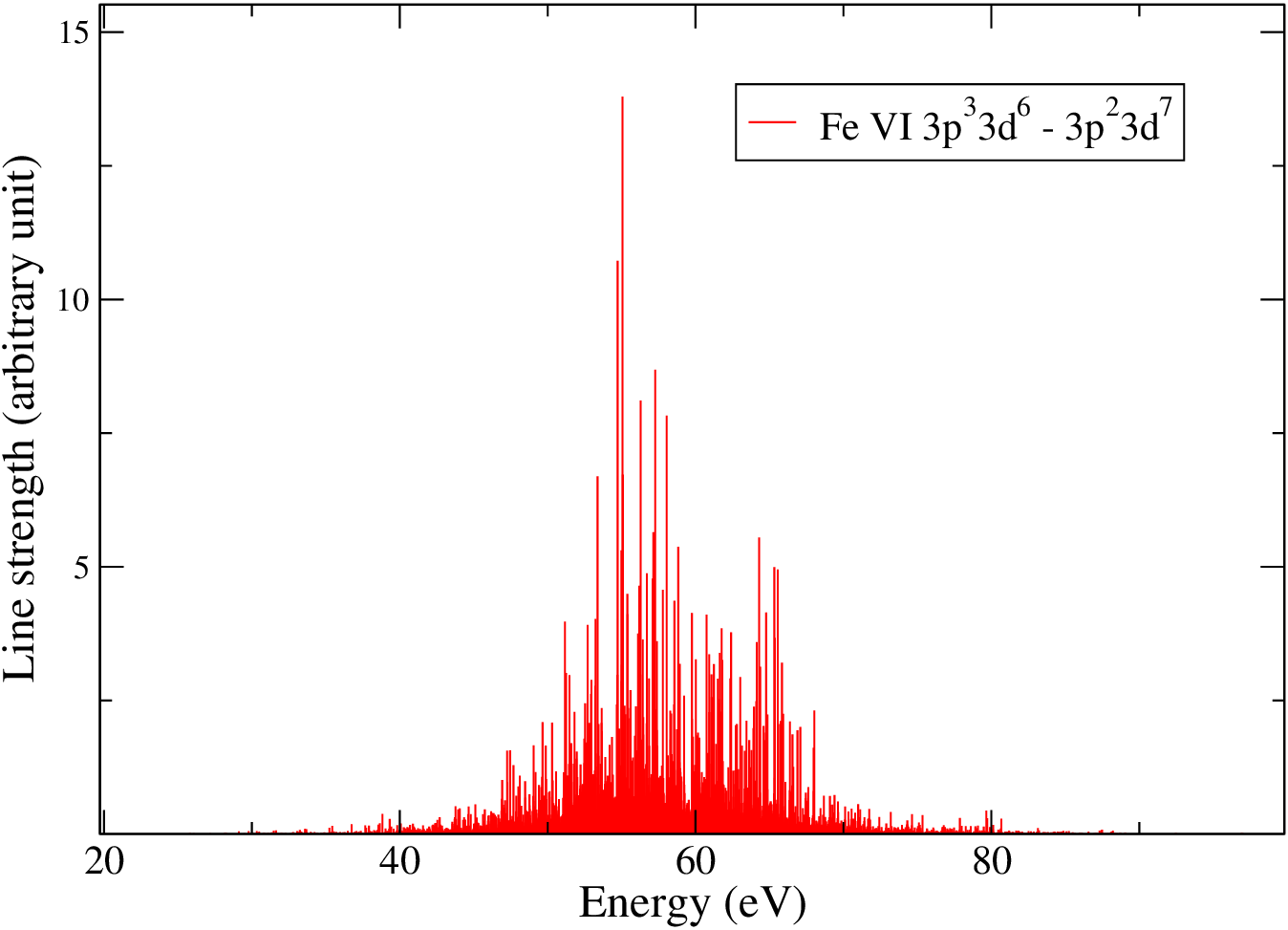}
\end{center}
\vspace{1cm}
\caption{(Color online) Line spectrum of the electric-dipole transition array Fe VI $3p^33d^6\rightarrow 3p^23d^7$.}\label{fig2}
\end{figure}

\begin{figure}
\begin{center}
\vspace{1cm}
\includegraphics[width=10cm]{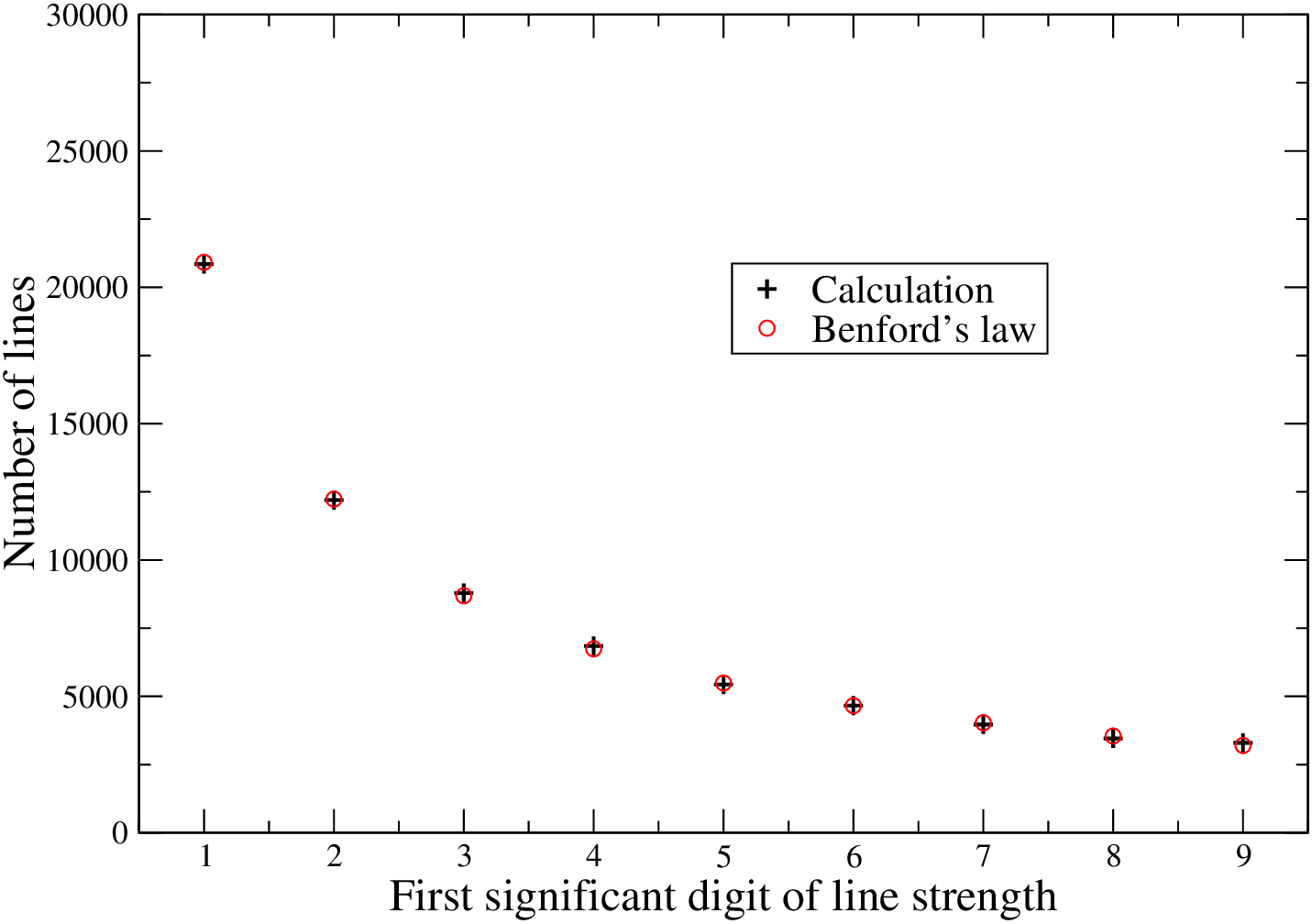}
\end{center}
\vspace{1cm}
\caption{(Color online) Significant digits of the lines of transition array Fe VI $3p^33d^6\rightarrow 3p^23d^7$.}\label{fig3}
\end{figure}

\begin{figure}
\begin{center}
\vspace{1cm}
\includegraphics[width=10cm]{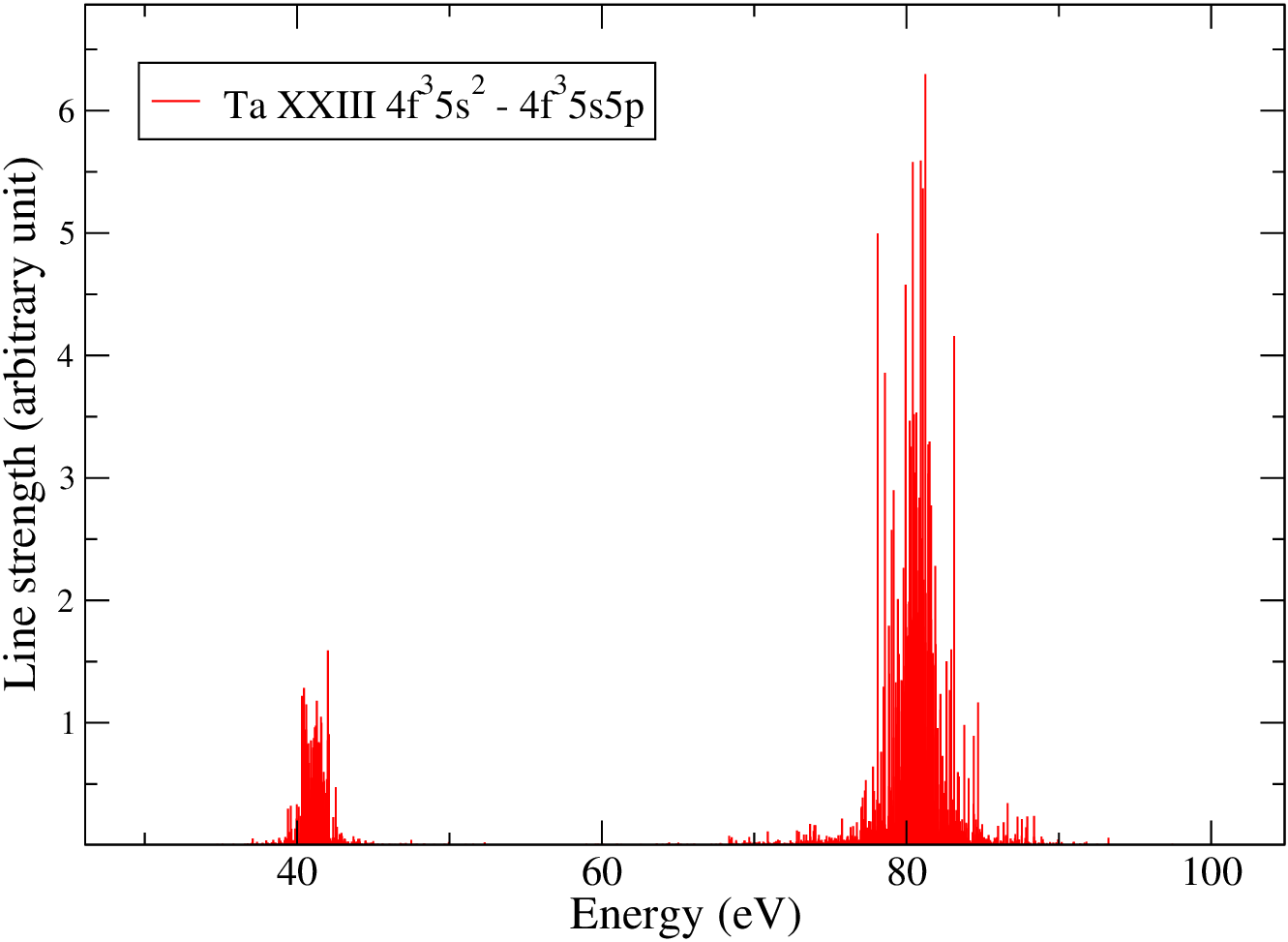}
\end{center}
\vspace{1cm}
\caption{(Color online) Line spectrum of the electric-dipole transition array Ta XXIII $4f^35s^2\rightarrow 4f^35s5p$.}\label{fig4}
\end{figure}

\begin{figure}
\begin{center}
\vspace{1cm}
\includegraphics[width=10cm]{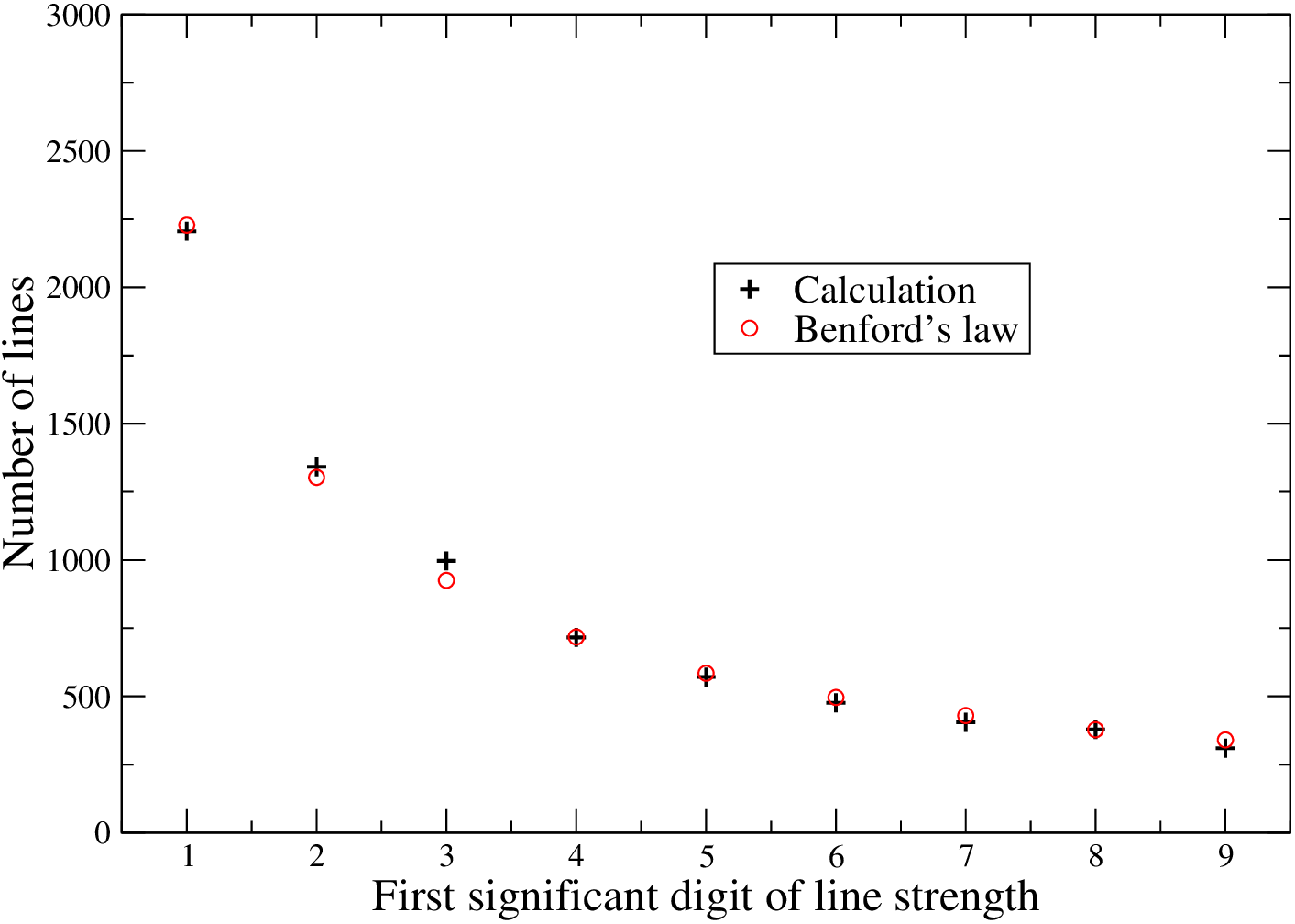}
\end{center}
\vspace{1cm}
\caption{(Color online) Significant digits of the lines of transition array Ta XXIII $4f^35s^2\rightarrow 4f^35s5p$.}\label{fig5}
\end{figure}

\noindent It was found recently (see Ref. \cite{PAIN08}) that the distribution of lines in a given transition array follows very well Benford's logarithmic law of significant digits - see Figs. \ref{fig2} and \ref{fig3}. This indicates that the distribution of digits reflects the symmetry due to the selection rules. If transitions were governed by uncorrelated random processes, each digit would be equi-probable. An interesting point is that the lines of a non-relativistic transition array seem to follow Benford's law even when the spin-orbit interaction is important, see Figs. \ref{fig4} and \ref{fig5}, \emph{i.e.}, when the array is split into relativistic subarrays.

\subsection{Explanations}\label{subsec32}

\subsubsection{Multiplicative stochastic processes}\label{subsec321}

Benford's law is still not fully understood mathematically. However, it can be proved to apply if the system is governed by random multiplicative processes \cite{PIETRONERO01}, \emph{i.e.}, processes which are additive in a logarithmic space. In Wigner's Random Matrix Theory (RMT), the Hamiltonian is defined in the GOE by an ensemble of real symmetric matrices whose probability distribution is a product of the distributions for the individual matrix elements  $H_{kl}$, considered as stochastic variables, and the variance of the distribution for the diagonal elements is twice the one for the off-diagonal elements. The matrix elements of the Hamiltonian are correlated stochastic variables and the product of such variables, arising through the diagonalization process, leads to Benford's logarithmic distribution of digits. 

Therefore, since Benford's law can be explained in terms of a dynamics governed by multiplicative stochastic processes, the RMT is probably an interesting pathway for the calculation of large atomic-dipole transition arrays  \cite{WILSON88} and Benford's law can help clarifying the existence of different classes of stochastic Gaussian variables.

\subsubsection{Scale invariance}\label{subsec322}

While the previous explanations show what kind of data would conform to Benford's law, scale invariance explains how the formula can be derived. If the first digits of some large data set conform to a particular distribution, then the latter distribution must be independent of the data's units of measurement. Let us consider that variable $x$ has a scale invariant distribution and assume that $1\leq x\leq 10$. If $x$ is scale invariant, multiplying $x$ by a constant, \emph{i.e.}, adding a constant to $\log_{10}(x)$ does not change the distribution. The only distribution invariant when a constant is added is the uniform distribution $U$. This means that in the interval $\left[1,10\right]$:

\begin{equation}
\log_{10}(x)\approx U\left[\log_{10}(1),\log_{10}(10)\right],
\end{equation}

\noindent and therefore $\hp(\log_{10}(x))=1$. Then, we have

\begin{eqnarray}
\hp(d=k)&=&\hp(k\leq x<k+1)\nonumber\\
&=&\hp\left(\log_{10}(k)\leq\log_{10}(x)<\log_{10}(k+1)\right)\nonumber\\
&=&\int_{\log_{10}(k)}^{\log_{10}(k+1)}dy=\log_{10}\left(1+\frac{1}{k}\right).
\end{eqnarray}

\begin{table}[t]
\begin{center}
\begin{tabular}{|c|c|}\hline
Line strength (arbitrary units) & First significant digit \\ \hline\hline
0.00001929 & 1 \\
0.00005922 & 5 \\
0.00162345 & 1 \\ 
0.00383879 & 3 \\ 
0.00062788 & 6 \\ 
0.00263967 & 2 \\ 
0.00000421 & 4 \\ 
0.00400046 & 4 \\ 
0.00022280 & 2 \\ 
0.00035794 & 3 \\ 
0.00000050 & 5 \\
0.00000054 & 5 \\ 
0.00000534 & 5 \\ 
0.00016042 & 1 \\ 
0.00000400 & 4 \\ 
0.00011443 & 1 \\
0.00022017 & 2 \\
0.00066048 & 6 \\
0.00022519 & 2 \\ 
0.00001100 & 1 \\ 
0.00067345 & 6 \\ 
0.00223977 & 2 \\\hline 
\end{tabular}
\end{center}
\caption{First significant digit of several line strengths of transition array Fe VI $3p^33d^6\rightarrow 3p^23d^7$.}\label{tab3}
\end{table}

\noindent The fact that the line strengths follow Benford's law, which is a consequence of scale invariance, the basic concept of fractals, is consistent with the fact that the distribution of the lines presents a fractal nature, as we have seen in Sec. \ref{sec2}.

\section{Propensity rule}\label{sec4}

The propensity rule belongs to a class of tendencies that are well understood by selection rules and correlation laws. A line connects preferentially a low (high) energy level of the initial configuration to a low (high) energy level of the final configuration. Such a correlation is due to the selection rules and to the fact that the energies of the levels of a configuration follow a preferential order with respect to the quantum numbers (\emph{cf.} Hund's rule for $\ell^N$ or $\ell^Ns$ configurations, which states that the levels of highest spin have the lowest energy). The selection rule $\Delta S$=0 for E1 lines imposes a peculiar kind of correlation between energies and amplitudes. Indeed, it was observed that   

\begin{equation}\label{cor}
\frac{\int_{-\infty}^{+\infty}dE_dE_d\int_{-\infty}^{+\infty}D\left(E_d,E_u,a\right)a^2da}{\int_{-\infty}^{+\infty}dE_d\int_{-\infty}^{+\infty}D\left(E_d,E_u,a\right)a^2da}\approx K . E_u
\end{equation}

\noindent where $D\left(E_d,E_u,a\right)$ represents the number of lines having an amplitude belonging to $\left[a,a+da\right]$ and connecting an energy level in $\left[E_d,E_d+dE_d\right]$ to an energy level in $\left[E_u,E_u+dE_u\right]$. Equation (\ref{cor}) illustrates the existence of a correlation coefficient $K$, which is equal to

\begin{equation}
K=\frac{\langle E_d.E_u\rangle}{\langle E_u^2\rangle},
\end{equation}

\noindent where the strength-weighted average value of quantity $X$ is defined as

\begin{equation}
\langle X\rangle=\frac{\sum_{du}XS_{du}}{\sum_{du}S_{du}}.
\end{equation}

\noindent The consequence of such a correlation is that the strongest lines are mostly located around the center of the transition array, for example, see Fig. \ref{fig2}.

\section{Concentration of the lines in the high-energy side: role of the $G^k$ exchange Slater integral}\label{sec5}

\begin{figure}
\begin{center}
\vspace{1cm}
\includegraphics[width=10cm]{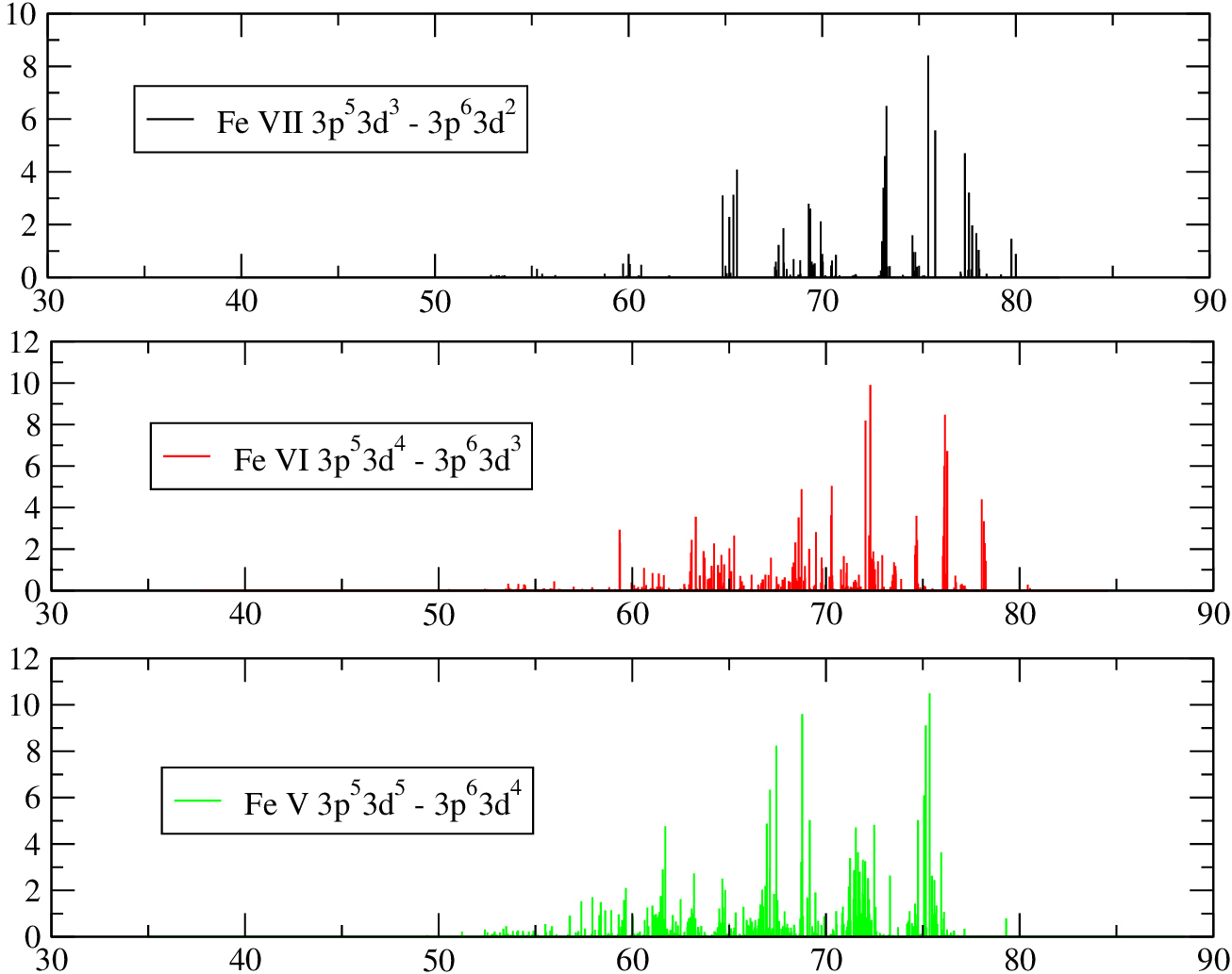}
\end{center}
\vspace{1cm}
\caption{(Color online) Evolution of transition array $3p^53d^Q\rightarrow 3p^63d^{Q-1}$ when $Q$ varies from 3 to 5. For $Q=3$, $\mathcal{C}\left[G^1\left(pd\right)\right]=133/135\approx 0.985$, for $Q=4$, $\mathcal{C}\left[G^1\left(pd\right)\right]=38/45\approx 0.845$ and for $Q=5$, $\mathcal{C}\left[G^1\left(pd\right)\right]=19/27\approx 0.704$.}\label{fig6}
\end{figure}

In practice the propensity rule is obviously not always satisfied, and one can observe a concentration of the oscillator strength towards the high-energy side of the transition array \cite{OSULLIVAN99}, see Fig. \ref{fig6}, which occurs as well as for some complex Auger spectra. For configurations $\ell^N\ell'^{N'+1}$ with two open subshells having the same principal quantum number, the Coulomb exchange interaction energy mainly determines the energy level spectrum. This interaction forms the upper and lower groups of levels with very different abilities to participate in transitions. Due to the relation between the position of a level and the transition amplitude from this level, the transitions mainly from the upper group of levels manifest themselves in the radiative or Auger spectra. The resulting asymmetrical shape of the transition array \cite{PAIN09,PORCHEROT11} is linked with the existence of the emissive zones, which are an emissivity-weighted energy level distribution of the upper configuration \cite{GILLERON11,BAUCHE83}, consisting in a difference between the energy of the center of gravity of the emissive zone and the energy of the upper configuration. In general, this goes hand in hand with a dominant exchange Slater integral $G^1$ with a positive coefficient, which is always the case in $\ell^{N+1}\rightarrow \ell^N\ell'$ arrays that are therefore always asymmetrical \cite{BAUCHE00}. It was proved by Racah that the inequality

\begin{equation}
\frac{G^k\left(\ell\ell'\right)}{2k+1}>\frac{G^{k+1}\left(\ell\ell'\right)}{2k+3}>0
\end{equation}

\noindent is always true whatever the value of $k$ \cite{RACAH42}, but it is worth mentioning that the exchange Slater integrals usually, although not always, satisfy the inequality

\begin{equation}
G^k\left(\ell\ell'\right)>G^{k+1}\left(\ell\ell'\right)>0.
\end{equation}

\noindent For the transition array $\ell^N\ell'^{N'+1}\rightarrow\ell^{N+1}\ell'^{N'}$, the shift of the emissive zone of $\ell^N\ell'^{N'+1}$ (\emph{i.e.}, the difference between the center of gravity of the emissive zone and the average energy of the upper configuration $\ell^N\ell'^{N'+1}$) reads \cite{GILLERON11}:

\begin{eqnarray}
\delta E&=&\frac{N\left(4\ell'+1-N'\right)}{\left(4\ell+1\right)\left(4\ell'+1\right)}\left(2\ell+1\right)\left(2\ell'+1\right)\nonumber\\
& &\times\left[\sum_k\left(\frac{2}{3}\delta_{k,1}-\frac{1}{2(2\ell+1)\left(2\ell'+1\right)}\right)\threej{\ell}{k}{\ell'}{0}{0}{0}^2G^k\left(\ell\ell'\right)\right.\nonumber\\
& &\left.+\sum_{k>0}\sixj{\ell}{\ell}{k}{\ell'}{\ell'}{1}\threej{\ell}{k}{\ell}{0}{0}{0}\threej{\ell'}{k}{\ell'}{0}{0}{0}F^k\left(\ell\ell'\right)\right],\nonumber\\
& &
\end{eqnarray}

\noindent where $F^k\left(\ell\ell'\right)$ are direct Slater integrals. The coefficient of the $G^1\left(\ell\ell'\right)$ exchange Slater integral is the only positive coefficient and is equal to

\begin{equation}
\mathcal{C}\left[G^1\left(\ell\ell'\right)\right]=\frac{N\left(4\ell'+1-N'\right)}{\left(4\ell+1\right)\left(4\ell'+1\right)}\left(\frac{2}{3}-\frac{1}{2\left(2\ell+1\right)\left(2\ell'+1\right)}\right)\ell_>,
\end{equation}

\noindent where $\ell_{>}=\max\left(\ell,\ell'\right)$. Values of $\mathcal{C}\left[G^1\left(df\right)\right]$ for $d^{N+1}f^{N'}\rightarrow d^Nf^{N'+1}$  are displayed in Table \ref{tab4}.

\begin{table}[t]
\begin{center}
\begin{tabular}{|c|c|}\hline
$d^{N+1}f^{N'}\rightarrow d^Nf^{N'+1}$ & $\mathcal{C}\left[G^1\left(df\right)\right]$ \\ \hline\hline
$d^7\rightarrow d^6f$ & $\frac{137}{105}\approx 1.305$ \\
$d^7f\rightarrow d^6f^2$ & $\frac{548}{455}\approx 1.204$ \\
$d^7f^2\rightarrow d^6f^3$ & $\frac{1507}{1365}\approx 1.104$ \\
$d^8\rightarrow d^7f$ & $\frac{137}{90}\approx 1.522$ \\
$d^8f\rightarrow d^7f^2$ & $\frac{274}{195}\approx 1.405$ \\
$d^8f^2\rightarrow d^7f^3$ & $\frac{1507}{1170}\approx 1.288$ \\
$d^9\rightarrow d^8f$ & $\frac{548}{315}\approx 1.740$ \\
$d^9f\rightarrow d^8f^2$ & $\frac{2192}{1365}\approx 1.606$ \\
$d^9f^2\rightarrow d^8f^3$ & $\frac{6028}{4095}\approx 1.472$ \\\hline 
\end{tabular}
\end{center}
\caption{Coefficient of the $G^1\left(df\right)$ exchange Slater integral in the shift of the emissive zone of $d^Nf^{N'+1}$.}\label{tab4}
\end{table}

In fact, the enhancement of some transitions and the near-forbidenness of the other transitions indicate the existence of an additional selection rule \cite{BERNOTAS01}. Let us consider the operator

\begin{equation}
A^{(10)}=\left[\tilde{b}^{(\ell s)\dagger}\times a^{(\ell's)\dagger}\right]^{(10)},
\end{equation}

\noindent where $a^{\left(\ell's\right)\dagger}$ is an electron creation operator with orbital rank $\ell$ and spin rank $s$. The operator $\tilde{b}^{\left(\ell s\right)\dagger}$ is a vacancy creation operator, equal to the annihilation operator $a^{\left(\ell s\right)}$, which only becomes the irreducible tensor operator after multiplication by a phase factor

\begin{equation}
\tilde{b}^{\left(\ell s\right)\dagger}_{m\mu}=(-1)^{\ell+s-m-\mu}~b^{\left(\ell s\right)\dagger}_{-m-\mu}=(-1)^{\ell+s-m-\mu}~a^{\left(\ell s\right)}_{-m-\mu}.
\end{equation}
 
\noindent The selection rule for the dipole transition amplitude follows from the relation between the operator $A^{(10)}$ and the radiative dipole transition operator $D^{(1)}$ in the second quantization representation for the considered transition $\ell^{N+1}\ell'^{N'}\rightarrow\ell^N\ell'^{N'+1}$:

\begin{equation}
D_q^{(1)}=\sqrt{\frac{2}{3}}A_{q0}^{(10)}\langle\ell'||C^{(1)}||\ell\rangle\langle n'\ell'|r|n\ell\rangle.
\end{equation}

\noindent The transition amplitude is \cite{BERNOTAS01}:

\begin{eqnarray}
& &\langle\ell^{N+1}\ell'^{N'}\gamma pLSM_LM_S|D_q^{(1)}|\ell^N\ell'^{N'+1}\gamma'p'L'SM_{L'}M_S\rangle\nonumber\\
&=&\delta\left(p,p'+1\right)\sqrt{\frac{2(N+1)N'}{3}}\sqrt{2L+1}\threej{1}{L'}{L}{q}{M_{L'}}{M_L}\nonumber\\
& &\times\left(\ell^N\ell'^{N'+1}\gamma' p'L'S',\ell^1\ell'^{4\ell'+1}~^1P\left\{|\right.\ell^{N+1}\ell'^{N'}\gamma pLS\right)\nonumber\\
& &\times\langle\ell'||C^{(1)}||\ell\rangle\langle n'\ell'|r|n\ell\rangle,
\end{eqnarray}

\noindent where $p$ is the number of vacancy-electron pairs. The selection rule is $p'=p-1$ and the small number of strong lines is related to the vicinity of the so-called PH (particle-hole) coupling, which corresponds to the case where the exchange integrals $G^k$ are assumed to be the only non-zero integrals. 

\section{Symmetry with respect to the quarter of a subshell ($LL$ coupling)}\label{sec6}

Regularities and tendencies are not exact features, but are often the visible part of exact symmetries. The symmetry with respect to a quarter of the subshell was discovered for the ratio of line intensities of the strongest lines emitted by lanthanides in calibrated arcs. The reason for that symmetry, which is satisfied by some important atomic quantities related to the ground state of atoms, is not obvious. It is well known that chemical properties of lanthanides and actinides are determined by the peculiarities of the inner atomic $f$ subshell. A configuration $\ell^N$ can be viewed as two subshells with spins of electrons directed up and down (``spin-polarized'' model):

\begin{equation}
\ell_{\downarrow}^{N'}\ell_{\uparrow}^{N-N'}.
\end{equation}
 
\noindent For the highest multiplicity, the decomposition is unique:

\begin{equation}
\ell^N\rightarrow\left\{\begin{array}{ll}
\ell_{\downarrow}^N & \mathrm{if} \;\;\; N\leq 2\ell+1\\
\ell_{\downarrow}^{2\ell+1}\ell_{\uparrow}^{N-2\ell-1} & \mathrm{if} \;\;\; N > 2\ell+1
                        \end{array},\right.
\end{equation}

\noindent and the electron-vacancy symmetry for the subshell having only $(2\ell+1)$ single-electron states manifests as a symmetry with respect to a quarter of the subshell:

\begin{equation}
\ell^N\rightarrow\left\{\begin{array}{ll}
\ell^{2\ell+1-N} & \mathrm{if} \;\;\; N\leq 2\ell+1\\
\ell^{6\ell+3-N} & \mathrm{if} \;\;\; N > 2\ell+1
                        \end{array}.\right.
\end{equation}

\begin{figure}
\begin{center}
\vspace{1cm}
\includegraphics[width=10cm]{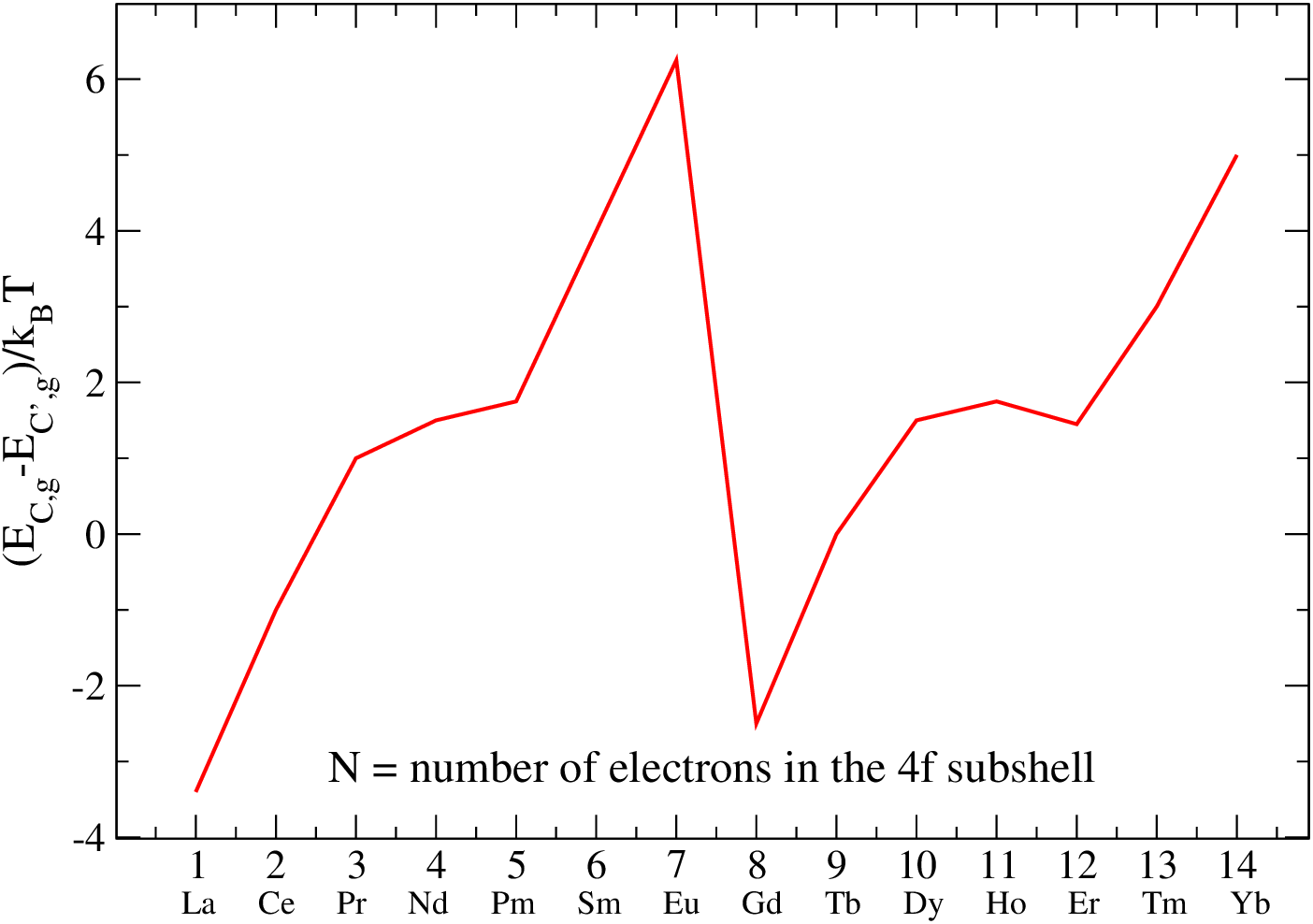}
\end{center}
\vspace{1cm}
\caption{(Color online) Difference between the groud state energies of the lowest levels of configurations $C=4f^{N-1}5d6s^2$ and $C'=4f^N6s^2$ versus the number of electrons in the $f$ subshell.}\label{fig7}
\end{figure}

\begin{figure}
\begin{center}
\vspace{1cm}
\includegraphics[width=10cm]{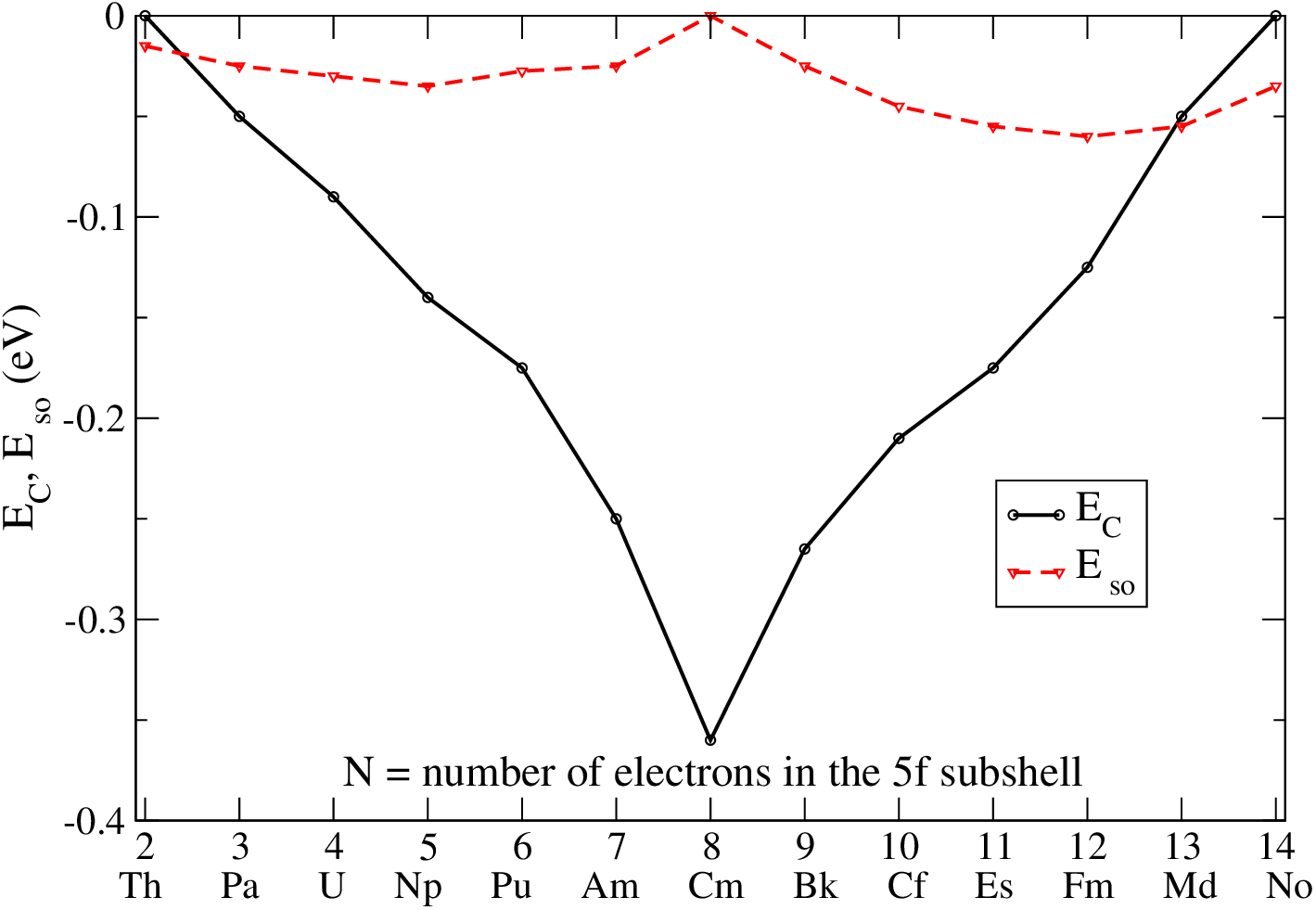}
\end{center}
\vspace{1cm}
\caption{(Color online) Variation of the Coulomb $E_C$ and spin-orbit $E_{\mathrm{so}}$ contributions to the energy of the ground states of actinides with respect to the number of electrons $N$ in the $f$ subshell .}\label{fig8}
\end{figure}

\noindent The signature of such a symmetry can be found in some physical quantities, for instance in the difference between the ground state energies of two configurations $C$ and $C'$, connected by the jump of a $f$ electron, versus the number of electrons in the $f$ subshell - see Fig. \ref{fig7}. It is also present in the variation of the spin-orbit part of ground-state energy of the configuration (but not in Coulomb part) of the $f$ subshell with respect to the number of electrons - see Fig. \ref{fig8}. 

\begin{figure}
\begin{center}
\vspace{1cm}
\includegraphics[width=10cm]{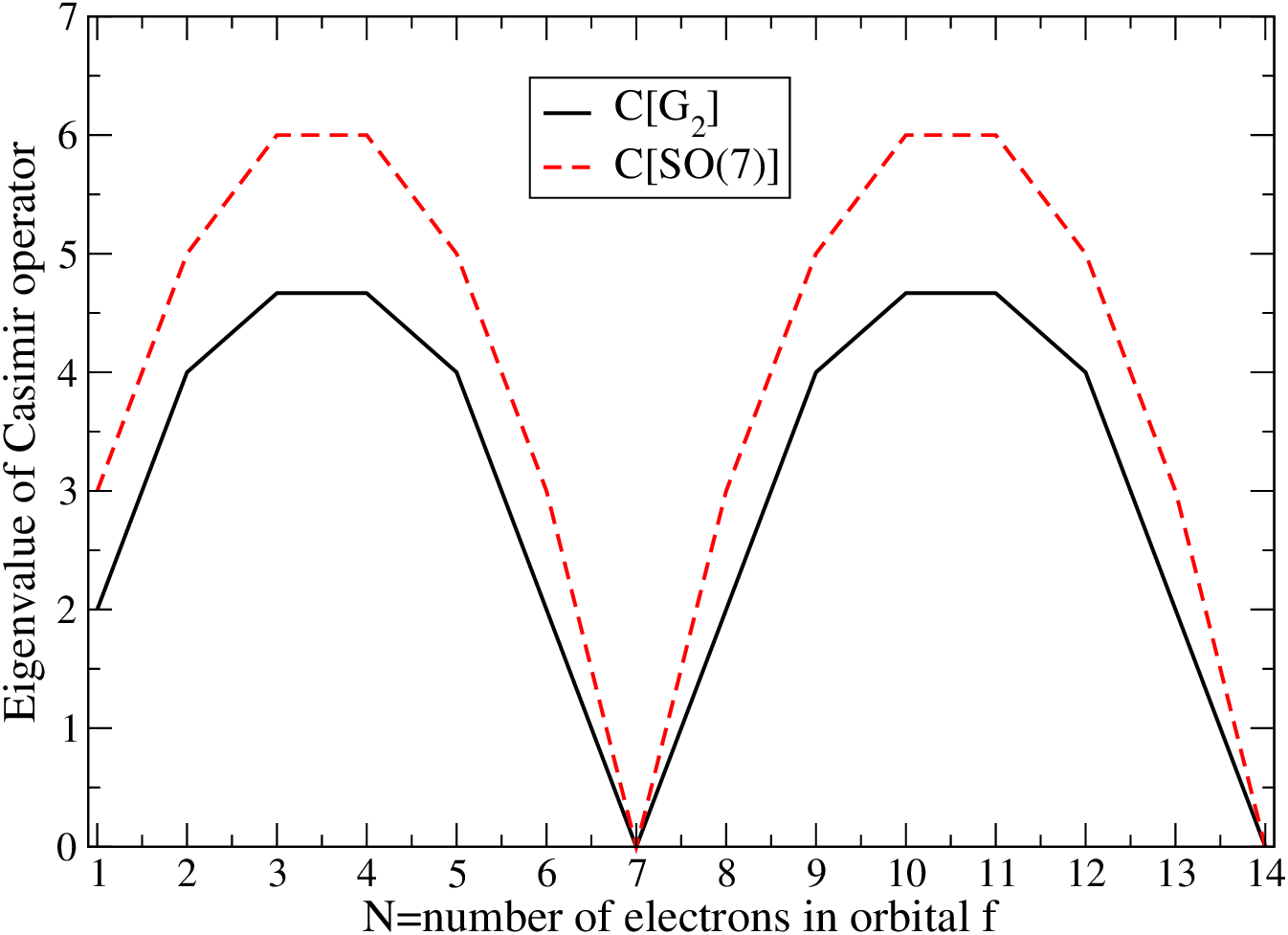}
\end{center}
\vspace{1cm}
\caption{(Color online) Eigenvalues of the Casimir operators of special groups $G_2$ and $SO(2\ell+1)$ for the ground state of the $f^N$ configuration. Important configuration-interaction effects are expressed in terms of these values.}\label{fig9}
\end{figure}

Transformations among the seven orbital states of an $f$ electron can be described by the unitary group $U(7)$ and its semi-simple Lie sub-groups $SO(7)$, $G_2$ and $SO(3)$. The symmetry with respect to a quarter of the subshell is not violated by configuration interaction effects, and the mixing leads to the correlation corrections \cite{RAJNAK63}

\begin{equation}
\Delta E_{\mathrm{cor}}=\alpha L(L+1)+\beta C\left[G_2\right]+\gamma C\left[SO(7)\right]
\end{equation}

\noindent where $C\left[G_2\right]$ and $C\left[SO(7)\right]$ are eigenvalues of the Casimir operators for the special groups $G_2$ and $SO(7)$ \cite{KARAZIJA96,KARAZIJA98}:

\begin{equation}
\left\{
\begin{array}{l}
C\left[G_2\right]=\frac{B(7-B)}{2^23^35}\left[6+41B(7-B)-2B^2(7-B)^2\right]\\
C\left[SO(7)\right]=\frac{B(7-B)}{2},
\end{array}
\right.
\end{equation}

\noindent where $B=N$ if $N\leq 7$ and $B=7-N$ otherwise. The symmetry with respect to a quarter of the subshell is clearly visible in $C\left[G_2\right]$ and $C\left[SO(7)\right]$ - see Fig. \ref{fig9}.

There are also interesting symmetry properties with respect to half of a subshell \cite{KARAZIJA08}. It was shown by Karazija and Momkauskait\.e that the following explicit formula for the fractional parentage coefficients for the ground term of a half-filled subshell $\ell^{2\ell+1}$ with $L=0$:

\begin{equation}
\left(\ell^{2\ell+1}L=0,\left(S=\ell+\frac{1}{2}\right)\left\{|\right.\ell^{2\ell}L'S'\ell\right)=\delta\left(L',\ell\right)\delta\left(S',\ell\right)
\end{equation}

\noindent provides the possibility of obtaining simple expressions for the intensities of the lines in photo-electron and photo-excitation spectra. 

\section{Example of asymmetry: the odd-even staggering in the parity of quantum number $J$}\label{sec7}

\noindent Statistical methods appear to be very useful for the modeling of complex atomic spectra. Indeed, approximate formulas have been derived for the values of the total atomic number $J$ in an electronic configuration, and for the number of electric-dipole lines in an atomic transition array. These problems have been solved exactly, by tedious explicit level-by-level methods \cite{CONDON35}, but no general formula exists. In addition, no fluctuations \emph{i.e.}, deviations of the exact results from the statistical formulae, have yet been characterized. Actually, one type of fluctuation is the predominance of the levels with even $J$ values. Studying the statistics of $M_J$ values consists in classifying the Slater determinants according to their values of $M_J$. Although the total number of determinants is equal to a simple combinatorial factor, no compact formula is yet known for the number of determinants relative to a given $M_J$ value. \noindent Let us consider a system of $N$ identical fermions in a configuration consisting of a single orbital of degeneracy $g$, $m_i$ being the angular momentum projection of electron state $i$. Two constraints must be satisfied:

\begin{equation}
N=n_1+\cdots+n_g=\sum_{i=1}^gn_i,
\end{equation}

\noindent where $n_i$ bis the number of electrons in state $i$ and

\begin{equation}
M_J=n_1m_1+\cdots+n_gm_g=\sum_{i=1}^gn_im_i,
\end{equation}

\noindent where $n_i=0$ or 1 $\forall i$. The generating function associated to that numbering problem reads

\begin{eqnarray}
f(x,z)&=&\sum_{N=0}^{\infty}\sum_{M_J=-\infty}^{\infty}z^Nx^{M_J}\nonumber\\
& &\times\sum_{\{n_1,\cdots,n_g\}}\delta_{N,n_1+\cdots+n_g}~.~\delta_{M_J,n_1m_1+\cdots+n_gm_g}\nonumber\\
&=&\sum_{\{n_1,\cdots,n_g\}}z^{n_1+\cdots+n_g}~.~x^{n_1m_1+\cdots+n_gm_g}.
\end{eqnarray}

\noindent Since the quantities $n_i$ are independent, it is possible to write

\begin{eqnarray}
f(x,z)&=&\sum_{n_1=0}^1z^{n_1}x^{n_1m_1}\cdots\sum_{n_g=0}^1z^{n_g}x^{n_gm_g}\nonumber\\
&=&\left(1+z~x^{m_1}\right)\times\cdots\times\left(1+z~x^{m_g}\right)=\prod_{i=1}^g\left(1+z~x^{m_i}\right).
\end{eqnarray}

\noindent In the particular case of the relativistic configuration $j^N$, the generating function associated to that problem is

\begin{equation}\label{eqfj}
f_j(x,z)=\prod_{m=-j}^j\left(1+x^mz\right)=\sum_{N=0}^{2j+1}z^Nf_{j,N}(x),
\end{equation}

\noindent The number of levels having angular momentum $J$ is given by the Condon-Shortley relation $Q(J)=P\left(M_J=J\right)-P\left(M_J=J+1\right)$, where $P\left(M_J=J\right)$ is the coefficient of $x^{M_J}$

\begin{equation}
f_{j,N}(x)=\sum_{M_J=M_{J,\mathrm{min}}}^{M_{J,\mathrm{max}}}P(M_J)~x^{M_J},
\end{equation}

\noindent with

\begin{equation}
M_{J,\mathrm{max}}=\sum_{m=j-N+1}^jm=N(2j+1-N)/2
\end{equation}

\noindent and $M_{J,\mathrm{min}}=-M_{J,\mathrm{max}}$ \cite{GILLERON09}. In the case of an even number of electrons ($N=2k$), the excess of $J$ values is equal to the excess of $M_J$ values \cite{BAUCHE97b} and therefore

\begin{eqnarray}
E\left(j^{2k}\right)&=&\sum_{M_J=M_{J,\mathrm{min}}}^{M_{J,\mathrm{max}}}P(M_J)~(-1)^{M_J}=f_{j,N}(-1)=\left.\frac{1}{(2k)!}\frac{\partial^{2k}}{\partial z^{2k}}f_j(-1,z)\right|_{z=0}\nonumber\\
&=&\bin{2j+1}{k}=\frac{(2j+1)!}{k!(2j+1-k)!}.
\end{eqnarray}

\noindent For configuration $\ell^{2k}$, Van der Monde identity yields

\begin{equation}
E\left(\ell^{2k}\right)=\underbrace{\sum_{i_1=0}^{\ell}\sum_{i_2=0}^{\ell+1}}_{i_1+i_2=k}\bin{\ell}{i_1}\bin{\ell+1}{i_2}=\bin{2\ell+1}{k}.
\end{equation}

\noindent For example, configuration $f^6$ contains 115 even levels and 80 odd levels; the excess is equal to 35. In a configuration $\ell_1^{N_1}\ell_2^{N_2}\ell_3^{N_3}\cdots\ell_w^{N_w}$ with $\sum_{i=1}^NN_i$ even, if two subshells have an odd number of electrons, \emph{e.g.}, for instance $2p^33d4d^2$, the number of even $J$ values is equal to the number of odd $J$ values , \emph{i.e.} the excess is equal to zero. In all the other situations, the number of even $J$ values is larger than the number of odd $J$ values, \emph{i.e.} the excess is positive, and equal to

\begin{equation}
E\left(\ell_1^{N_1}\ell_2^{N_2}\ell_3^{N_3}\cdots \ell_w^{N_w}\right)=\prod_{i=1}^w\bin{2\ell_i+1}{N_i/2}.
\end{equation}

\noindent For instance, the excess of configuration $2p^43d^64f^2$ is equal to 210. It appears that the relative importance of the excess is a rapidly decreasing function of the complexity of the configurations. Bauche and Coss\'e checked the excess in published tables of atomic energy levels \cite{SUGAR85}, using the first series of transition elements and discarding the levels liying above the first ionization limit, as they have been classified by methods other than the Ritz combination principle.

\section{Conclusion}\label{sec8} 

Searching for regularities and trends in atomic spectroscopy can be of great interest, in order to better understand the physical processes underlying the structure and spectra of complex atomic systems. They can be helpful in order to check the reliability of theoretical predictions as well as experimental measurements. The atomic levels discovered by physicists thriving in the classification of line spectra constitute a priceless material for testing the fundamental laws of statistics. We found that Learner's distribution of electric-dipole lines possesses a fractal character, and that its dimension is close to $1/3$. The rule remains a mystery, but can be related to other observed peculiarities of spectra such as Benford's law \cite{PAIN08} for the distribution of strengths, which can be explained by scale invariance. Actually, the natural development of fractal analysis of sets with a complicated structure is their description as multifractals \cite{FEDER88}. A multifractal represents a set of elementary fractals that are organized by means of some measure, which characterizes the density of the physical quantity on its geometrical carrier. Such mathematical tools should bring useful additional informations. It would also be worth investigating the case of other multipolarities (electric or magnetic) \cite{PAIN12a} or to study the eventual fractal character of anomalous Zeeman patterns \cite{PAIN12b}. Some other properties were also discussed, such as the propensity law, the existence of a selection rule for the number of vacancy-electron pairs (PH coupling) and the symmetry with respect to a quarter of the subshell, which consequences on the line-intensity ratios is still not fully understood, in the spin-adapted space (LL coupling), emphasizing the crucial role of group theory in atomic spectroscopy. Finally, an example of asymmetry, the odd-even staggering in the values of the total atomic angular momentum, was presented using the generating-function technique.

\section*{Acknowledgements}

The authors would like to thank J. Bauche for raising some of the issues presented here to his attention and P. Lesaffre for helpful discussion about the fractal dimension.


\begin{thebibliography}{99}

\bibitem{JOHANSSON96} S. Johansson, Phys. Scr. {\bf T65}, 7 (1996).

\bibitem{KUHN62} H. G. Kuhn, {\it Atomic Spectra} (Longmans, London, 1962).

\bibitem{RACAH55} G. Racah, in ``Proceedings of the Rydberg Centennial Conference on Atomic Spectroscopy'' (Edited by B. Edl\'en), Lunds Universitet \AA rsskrift, N. F. Avd. 2, Bd. 50 (C. W. K. Gleerup, Lund 1955) p. 31.

\bibitem{JUDD85} B. R. Judd, Rep. Prog. Phys. {\bf 48}, 907 (1985).

\bibitem{FLAMBAUM94} V. V. Flambaum, A. A. Gribakina, G. F. Gribakin and M. G. Kozlov, Phys. Rev. A {\bf 50}, 267 (1994).

\bibitem{FLAMBAUM98} V. V. Flambaum, A. A. Gribakina and G. F. Gribakin, Phys. Rev. A {\bf 58}, 230 (1998).

\bibitem{CONNERADE98} J.-P. Connerade, {\it Highly excited atoms} (Cambridge University Press, Cambridge, 1998).

\bibitem{BAUCHE79} C. Bauche-Arnoult, J. Bauche and M. Klapisch, Phys. Rev. A {\bf 20} 2424 (1979).

\bibitem{COWAN81} R. D. Cowan, \textit{The theory of atomic structure and spectra} (University of California Press, Berkeley and Los Angeles, 1981).

\bibitem{SARANDAEV95} E. V. Sarandaev, M. Kh. Salakhov and I. S. Fishman, J. Quant. Spectrosc. Radiat. Transfer {\bf 54}, 651 (1995).

\bibitem{BETHE57} H. A. Bethe and E. E. Salpeter, {\it Quantum Mechanics of One- and Two-Electron Atoms} (Springer, Berlin, 1957).

\bibitem{LEARNER82} R. C. M. Learner, J. Phys. B \textbf{15}, L891 (1982).

\bibitem{SCHEELINE86a} A. Scheeline, Anal. Chem. {\bf 58}, 802 (1986).

\bibitem{SCHEELINE86b} A. Scheeline, Anal. Chem. {\bf 58}, 3103 (1986).

\bibitem{HOWARD85} L. E. Howard and K. L. Andrew, J. Opt. Soc. Am. B {\bf 2}, 1032 (1977). 

\bibitem{BAUCHE97a} C. Bauche-Arnoult and J. Bauche, J. Quant. Spectrosc. Radiat. Transfer {\bf 58}, 441 (1997).

\bibitem{KURUCZ} R. L. Kurucz, {\it Atomic Line List}, CD-ROM No. 23. Harvard-Smithsonian Center for Astrophysics.

\bibitem{MANDELBROT82} B. B. Mandelbrot, {\it The Fractal Geometry of Nature} (Freeman, San Francisco, 1982). 

\bibitem{CEDERBAUM85} L. S. Cederbaum, E. Haller and P. Pfeifer, Phys. Rev. A {\bf 31}, 1869 (1985).

\bibitem{ZIMMERMANN87} Th. Zimmermann, L. S. Cederbaum, H.-D. Meyer and H. K\"oppel, J. Phys. Chem. {\bf 91}, 4446 (1987).

\bibitem{ZIMMERMANN88} Th. Zimmermann, H. K\"oppel, L. S. Cederbaum, G. Persch and M. Demtr\"oder, Phys. Rev. Lett. {\bf 61}, 3 (1988).

\bibitem{WANG97} W. F. Wang and P. P. Ong, Phys. Rev. A {\bf 55}, 1522 (1997).

\bibitem{DEVITO88} C. L. DeVito and W. A. Little, Phys. Rev. A {\bf 38}, 6362 (1988).

\bibitem{LEVY34} P. L\'evy, C. R. Acad. Sci. Paris {\bf 198}, 424 (1934).

\bibitem{BAUCHE90} J. Bauche and C. Bauche-Arnoult, Comput. Phys. Rep. \textbf{12}, 1 (1990).

\bibitem{KISTENEV01} Yu. V. Kistenev and Yu. N. Ponomarev, Opt. Spectrosc. {\bf 90}, 362 (2001), translated from Optika i Spektroskopiya {\bf 90}, 419 (2001).

\bibitem{ROSENZWEIG60} N. Rosenzweig and C. E. Porter, Phys. Rev. {\bf 120}, 1698 (1960).

\bibitem{PORTER56} C. E. Porter and R. G. Thomas, Phys. Rev. {\bf 104}, 483 (1956).

\bibitem{PORTER65} C. E. Porter, {\it Statistical Theories of Spectra: Fluctuations}, Academic Press, New York, NY (1965).

\bibitem{GRIMES83} S. M. Grimes, Phys. Rev. C {\bf 28}, 471 (1983).

\bibitem{BAUCHE91} J. Bauche, C. Bauche-Arnoult, J.-F. Wyart, P. Duffy and M. Klapisch, Phys. Rev. A {\bf 44}, 5707 (1991).

\bibitem{BISSON91} S. E. Bisson, E. F. Worden, J. G. Conway, B. Comaskey, J. A. D. Stockdale and F. Nehring, J. Opt. Soc. Am. B {\bf 8}, 1545 (1991).

\bibitem{BOULIGAND29} G. Bouligand, Bull. Sci. Math. {\bf 2}, 185 (1929).

\bibitem{KOLMOGOROV58} A. N. Kolmogorov, Dokl. Akad. Nauk SSSR {\bf 119}, 861 (1958).

\bibitem{NEWCOMB81} S. Newcomb, Am. J. Math. \textbf{4}, 39 (1881).

\bibitem{BENFORD38} F. Benford, Proc. Am. Philos. Soc. \textbf{78}, 551 (1938).

\bibitem{PAIN08} J.-C. Pain, Phys. Rev. E {\bf 77}, 012102 (2008).

\bibitem{PIETRONERO01} L. Pietronero, E. Tossati, V. Tossati and A. Vespignani, Physica A \textbf{293}, 551 (2001).

\bibitem{WILSON88}  B. G. Wilson, F. Rogers and C. Iglesias, Phys. Rev. A {\bf 37}, 2695 (1988).

\bibitem{OSULLIVAN99} G. O'Sullivan, P. K. Carroll, P. Dunne R. Faulkner, C. McGuinness and N. Murphy, J. Phys. B: At. Mol. Opt. Phys. {\bf 32}, 1893 (1999).

\bibitem{PAIN09} J.-Ch. Pain, F. Gilleron, J. Bauche and C. Bauche-Arnoult, High Energy Density Phys. {\bf 5}, 294 (2009).

\bibitem{PORCHEROT11} Q. Porcherot, J.-Ch. Pain, F. Gilleron and T. Blenski, High Energy Density Phys. {\bf 7}, 234 (2011).

\bibitem{GILLERON11} F. Gilleron, J.-Ch. Pain, Q. Porcherot, J. Bauche and C. Bauche-Arnoult, High Energy Density Phys. {\bf 7}, 277 (2011).

\bibitem{BAUCHE83} J. Bauche, C. Bauche-Arnoult, E. Luc-Koenig, J.-F. Wyart and M. Klapisch, Phys. Rev. A {\bf 28}, 829 (1983).

\bibitem{BAUCHE00} C. Bauche-Arnoult, J. Bauche, J.-F. Wyart and K. B. Fournier, J. Quant. Spectrosc. Radiat. Transfer {\bf 65}, 57 (2000).

\bibitem{RACAH42} G. Racah, Phys. Rev. {\bf 62}, 460 (1942).

\bibitem{BERNOTAS01} A. Bernotas and R. Karazija, J. Phys. B: At. Mol. Opt. Phys. {\bf 34}, L741 (2001).

\bibitem{RAJNAK63} K. Rajnak and B. G. Wybourne, Phys. Rev. {\bf 132}, 280 (1963).

\bibitem{KARAZIJA96} R. Karazija, A. Udris, A. Kynien\.e and S. Ku\u{c}as, J. Phys. B: At. Mol. Opt. Phys. {\bf 29}, L405 (1996)

\bibitem{KARAZIJA98} R. Karazija and A. Kynien\.{e}, J. Phys. Chem. A {\bf 102}, 897 (1998).

\bibitem{KARAZIJA08} R. Karazija and A. Momkauskait\.e, Phys. Scr. {\bf 78}, 065301 (2008).

\bibitem{CONDON35} E. U. Condon and G. H. Shortley, {\it The theory of atomic spectra} (Cambridge, England: Cambridge University, 1935).

\bibitem{SUGAR85} J. Sugar and C. Corliss, {\it Atomic Energy Levels of the Iron-Period Elements: Potassium through Nickel}, National Bureau of Standards (Washington D. C.: US Govt Printing Office).\begin{scriptsize}\begin{tiny}\end{tiny}\end{scriptsize}

\bibitem{GILLERON09} F. Gilleron and J.-C. Pain, High Energy Density Phys. {\bf 5}, 320 (2009).

\bibitem{BAUCHE97b} J. Bauche and P. Coss\'e, J. Phys. B: At. Mol. Opt. Phys. {\bf 30}, 1411 (1997).

\bibitem{FEDER88} J. Feder, {\it Fractals} (Plenum, New York, 1988).

\bibitem{PAIN12a} J. C. Pain, F. Gilleron, J. Bauche and C. Bauche-Arnoult, J. Phys. B: At. Mol. Opt. Phys. {\bf 45}, 135006 (2012).

\bibitem{PAIN12b} J.-C. Pain and F. Gilleron, Phys. Rev. A {\bf 85}, 033409 (2012).

\end{thebibliography}
\end{document}